\documentclass[english,twocolumn,aps,prb,superscriptaddress,showpacs,floatfix]{revtex4-1}

\usepackage{amsmath}
\usepackage{graphicx}%,subfigure}% Include figure files
\usepackage{dcolumn}% Align table columns on decimal point
\usepackage{bm}% bold math
\usepackage{mathrsfs}
\draft % marks overfull lines with a black rule on the right

\begin{document}

\title{Coarse-Grain Model for Lipid Bilayer Self-Assembly and Dynamics: \\Multiparticle Collision Description of the Solvent} %Title of paper

\author{Mu-Jie Huang}
% \email{mjthebes@gmail.com.}
\affiliation{
Department of Physics, National Central University, Jhongli 32001, Taiwan}
%\email[]{Your e-mail address}

%\homepage[]{Your web page}

%\thanks{}

%\altaffiliation{}

\author{Raymond Kapral}
\affiliation{
Chemical Physics Theory Group, Department of Chemistry, University of Toronto, Toronto, Ontario M5S 3H6, Canada}

\author{Alexander S. Mikhailov}
\affiliation{
Abteilung Physikalische Chemie, Fritz-Haber-Institut der Max-Planck-Gesellschaft, Faradayweg 4-6, 14195 Berlin, Germany
}
\author{Hsuan-Yi Chen}
\affiliation{
Department of Physics, National Central University, Jhongli 32001, Taiwan}
\affiliation{
Institute of Physics, Academia Sinica, Taipei 11520, Taiwan
}
\affiliation{
Physics Division, National Center for Theoretical Sciences, Hsinchu, 30013, Taiwan
}
% Collaboration name, if desired (requires use of superscriptaddress option in \documentclass).

% \noaffiliation is required (may also be used with the \author command).

%\collaboration{}

%\noaffiliation

\date{\today}

\begin{abstract}
A mesoscopic coarse-grain model for computationally-efficient simulations of biomembranes is presented. It combines molecular dynamics simulations for the lipids, modeled as elastic chains of beads, with multiparticle collision dynamics for the solvent. Self-assembly of a membrane from a uniform mixture of lipids is observed. Simulations at different temperatures demonstrate that it reproduces the gel and liquid phases of lipid bilayers. Investigations of lipid diffusion in different phases reveals a crossover from subdiffusion to normal diffusion at long times. Macroscopic membrane properties, such as stretching and bending elastic moduli, are determined directly from the mesoscopic simulations.  Velocity correlation functions for membrane flows are determined and analyzed.

\end{abstract}

% insert suggested PACS numbers in braces on next line

\maketitle %\maketitle must follow title, authors, abstract and \pacs

% Body of paper goes here. Use proper sectioning commands.

% References should be done using the \cite, \ref, and \label commands
\section{Introduction}

Biological membranes, formed by lipid bilayers, play a fundamental role in the function of biological cells and the theoretical description of their structure, properties and dynamics is an important and challenging problem~\cite{gennis89,MBC94}. While powerful analytical theories exist for such systems~\cite{nelson04,safran94}, they treat membranes as mathematical surfaces and do not reproduce the physical bilayer structure. Therefore, they are applicable only on scales that are much larger than the actual membrane thickness. Moreover, they include phenomenological parameters which still need to be determined either from experiments or from microscopic simulations. For biological processes inside a cell, the relevant length scales lie in the nanometer and submicrometer ranges. In order to consistently describe processes involving micro-vesicles, membrane proteins and ion channels, theory and simulation that account for the lipid structure of a bilayer are required.

All-atom molecular dynamics (MD) simulations of lipid bilayers have been performed~(see, e.g.,~Ref.~[\onlinecite{tieleman97,feller2000,Groot_Grubmuller_01,saiz_klein_02,dickey08}]). However, they are costly and limited to relatively small systems and short time scales. While short-time-scale simulations are sufficient for the exploration of some aspects of membrane dynamics, there are many important biochemical processes which occur on longer time scales. For instance, it is known that characteristic mechanochemical motions in proteins, essential for their enzyme or motor functions, usually require milliseconds or more for their completion\cite{MBC94}. Hence, microscopic investigations of biomembranes with protein inclusions are beyond the capacity of all-atom MD simulations. Furthermore, such simulations are also too slow to microscopically reproduce the self-assembly of vesicles, structural instabilities of membranes or the effects of slow hydrodynamical modes on membrane dynamics.

This has prompted the development of a variety of coarse-grain simulation methods for biomembranes, which are still able to resolve important aspects of the lipid bilayer structure~\cite{Nielsenetal04,Goetz_Gompper_Lipowsky_99,lyubartsev05,venturolia06,V08,bennun09,orsi10,Lipowsky_04}.
Typically, a lipid molecule is modeled as a chain comprising one hydrophilic and several hydrophobic beads connected by elastic springs; each of these beads corresponds to a certain atomic group. In coarse-grain solvent descriptions, the solvent molecules are also represented by groups of atoms.

In implicit solvent models\cite{Farago_03,brannigan05,cook05,Reynwar_Illya_Harmandaris_muller_kermer_Deserno_07,deserno09}, the solvent particles are not actually included in a simulation and hydrophobic effects due to the presence of such particles are taken into account through the use of a tunable interaction potential between the lipids. Such a simplification results in a computational speed-up, making simulations of large-scale membrane instabilities possible~\cite{Reynwar_Illya_Harmandaris_muller_kermer_Deserno_07}. However, in such solvent-free models the coupling of biomembranes to hydrodynamic flows, as well as the hydrodynamic interactions mediated by the solvent, cannot be described.

In explicit solvent models employing dissipative particle dynamics (DPD), the solvent particles are included into the dynamical description, but actual molecular interactions between them are replaced by effective soft interaction potentials, so that the particles are allowed to penetrate one another. The use of a soft-core potential for the solvent and lipids makes it possible to employ much larger molecular dynamics integration time steps compared with those in all-atom  MD simulations; therefore, substantially accelerating the computation~\cite{V08,Shillcock_Lipowshy_02,Laradji_Kumar_04,Gao_Shillcock_Lipowsky_07}. Nonetheless, further acceleration is desirable.

A major portion of the computational time in explicit solvent models is spent simulating the dynamics of the large number of solvent molecules in the system. This suggests that it is desirable to construct a coarse-grain dynamical scheme that treats the solvent part of the dynamics efficiently.
Such a scheme is provided by multiparticle collision (MPC) dynamics~\cite{Malevantes_Kapral_99,Malevantes_Kapral_00}. In this approach, solvent particles, representing coarse-grained real molecules, free stream and undergo effective multiparticle collisions at discrete time moments. The collision and streaming rules are formulated in such a way that the mass and momentum conservation laws are satisfied. These rules can be constructed so that the dynamics is either micro-canonical and preserves the phase-space volume or is canonical at constant temperature. MPC dynamics has been applied to a variety of problems where fluid micro-flows were essential and there were interactions between fluids and macromolecules. Reviews of this method are available~\cite{Kapral_08,Gompper_Ihle_Kroll_Winkler_09}.

MPC dynamics has already been used for simulations of biomembranes. This method has been employed\cite{Noguchi_Gommpper_05} to study a micron-size vesicle under shear flow.   In this work, the membrane was modeled as a triangulated surface described by vertices connected by tethers; the lipid bilayer structure was not resolved.  In another study, a special color-collision rule was used to account for the interaction between the MCP solvent and the coarse-grained lipids\cite{Inoue_Takagi_Matsumoto_08}.  In our investigation a coarse-grain description of the lipid bilayer, resolving membrane structure, is combined with MPC dynamics for the solvent. Interactions between lipids and solvent particles are explicitly taken into account.

In Sec.~\ref{sec:Model}, the detailed formulation of the simulation method is given. Simulations for membranes at three different temperatures are presented in Sec.~\ref{sec:Membrane Properties}. The simulations can reproduce a gel phase at low temperature and a liquid phase at higher temperatures. Density profiles for lipid particles across the membrane, lipid chain order parameters, and radial distribution functions of lipid head particles are determined and discussed. Through direct simulations, intra-membrane diffusion is explored and a subdiffusion regime on relatively short time scales is observed. In the next sections, our investigations focus on the membrane in the liquid phase, important for biological applications. In Sec.~\ref{sec:self_assembly}, self-assembly of a membrane from an initially uniform mixture of lipids is demonstrated. The surface tension coefficient is determined from simulations on membranes of different sizes in Sec.~\ref{sec:Statistical_properties}. By constructing and analyzing the power spectrum of membrane height fluctuations, the elastic bending modulus of the membrane is found, fluctuations of the membrane flow velocity are considered and velocity-velocity correlation functions are analyzed. The paper ends with conclusions and a discussion of the results.

\section{Mesoscopic Model for Lipid Bilayer Dynamics\label{sec:Model}}

In this section we describe the mesoscopic coarse-grain model for the structure and dynamics of a lipid bilayer membrane in a solvent. The mesoscopic model uses a coarse-grain description of a lipid molecule as a collection of linked molecular groups termed beads. In addition, the solvent in which the lipids reside is treated at a particle-based level where each effective point solvent particle represents a collection of real solvent molecules. The coarse-grained lipid molecules interact through intermolecular potentials. The solvent particles also interact with the lipid beads through intermolecular potentials; however, the solvent particles interact among themselves through multiparticle collisions. There are no intermolecular interactions among solvent particles. The dynamical evolution of the entire systems, lipids plus solvent, is described by a hybrid dynamical scheme that combines molecular dynamics for all interacting particles with multiparticle collision dynamics for the solvent. The fact that there are no explicit solvent-solvent molecule interactions is responsible for the computational efficiency of this dynamical scheme. Below we provide a detailed description of the mesoscopic MD-MPC dynamical bilayer model.

%%%%%%%%%%%%%%%%%%%%%%%
\subsection{Lipid interactions}
\begin{figure}
\includegraphics[width=.8\columnwidth]{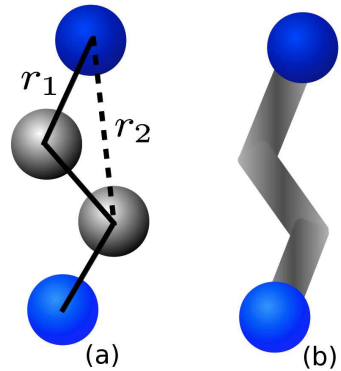}
\caption{\label{fig:lipid} The lipid chain (left) and its schematic representation as a rod (right). The lipid consists of four beads linked by  elastic FENE bonds (solid lines) and straightened by elastic bonds (dashed lines). The first bead (dark gray/blue) is hydrophilic. Three other beads are hydrophobic, the terminal hydrophobic bead is shown as light gray/blue}.
\end{figure}

A lipid chain comprises a hydrophilic head and a hydrophobic tail. In common with many other coarse-grain descriptions, a lipid molecule is modeled as a set of beads. In our investigation, we adopt a four-bead representation of the lipid where the hydrophobic head (h) is modeled as a single bead and the hydrophobic tail (t) as three beads (see Fig.~\ref{fig:lipid}(a)). Below, we specify the interactions between the beads in a lipid and between the lipids. These lipid interaction potentials have the same forms as in Cooke, et al.~\cite{Cooke_Kremer_Deserno_05}.

The interaction between two lipid beads is described by the truncated Lennard-Jones (LJ) potential,
\begin{equation}
V_{rep}(r_{ij}) =
4\epsilon_{\alpha \alpha'}\bigg[\bigg(\frac{\sigma}{r_{ij}}\bigg)^{12}-\bigg(\frac{\sigma}{r_{ij}}\bigg)^6 + \frac{1}{4}\bigg]\mbox{ } \theta(r_c-r_{ij}),
\label{repulsive_potential}
\end{equation}
where $\theta(r)$ is the Heaviside function and $r_{ij}=|\mathbf{r}_i-\mathbf{r}_j|$ is the distance between the beads $i$ and $j$.  The cutoff length $r_c=2^{1/6}\sigma$ is chosen in such a way that there is a short-distance repulsion but the long-distance attraction is absent. The strength of the interaction between beads $i$ and $j$ takes the value $\epsilon_{\alpha \alpha'}$, where $\alpha, \alpha' \in \{h,t\}$ if bead $i$ is of type $\alpha$ and bead $j$ is of type $\alpha'$.

Two neighboring particles in a lipid chain are linked by a FENE bond~\cite{Kremer_Grest_90}, described by the potential
\begin{equation}
V_{bond}=-\frac{1}{2}k_{bond} \ r_{\infty}^2\ln[1-(r_1/r_{\infty})^2]\mbox{ },
\end{equation}
where $r_1$ is the distance between the beads and $r_{\infty}=1.5\:\sigma$ is the maximum distance allowed by the FENE bond.  In all simulations, we have chosen the spring constant as $k_{bond}=20\:\epsilon/\sigma^2$, so that at equilibrium the length of the FENE bond is close to $\sigma$. Bending rigidity of a lipid chain is modeled by introducing additional springs connecting next-nearest neighbor beads and is described by the bending potential,
\begin{equation}
V_{bend}=\frac{1}{2}k_{bend}(r_2-4\sigma)^2.
\end{equation}
where $r_2$ is the distance between such two beads. The spring constant is $k_{bend}=2.5\:\epsilon/\sigma^2$ and the natural length is $4\:\sigma$. For a slightly bent lipid chain with FENE bond length $\sigma$, this potential reduces to $\frac{1}{2}k_{bend}\sigma^2\theta^2$,
where $\pi-\theta$ is the angle between two neighboring FENE bonds. Hence, it  provides a bending stiffness of $2.5 \epsilon$ to the lipid.

Hydrophobic effects are responsible for the aggregation of lipids into a membrane. These were taken into account by adding an attractive potential between beads that belong to different lipid tails.  The effective interaction was chosen to be

\begin{equation}
	V_{att}(r_{ij}) = \left\{
	\begin{array}{ll}
	4\epsilon_{\alpha\alpha'}\bigg[\big(\frac{\sigma}{r_{ij}}\big)^{12}-\big(\frac{\sigma}{r_{ij}}\big)^6 \bigg]\mbox{ },& r_{ij}< r_c\\
	%-\epsilon_{\alpha\alpha'},& r_{ij}< r_c\\
	\\
	-\epsilon_{\alpha\alpha'}\cos^2{\frac{\pi (r_{ij}-r_c)}{2w_{\alpha\alpha'} }}\mbox{ },& r_c\leq r_{ij}\leq r_c + w_{\alpha\alpha'}\\
	\\
	0\mbox{ },& r_{ij} > r_c + w_{\alpha\alpha'},
	\end{array}
	\right.
	 \label{attarctive_potential}
\end{equation}
where $w_{\alpha\alpha'}=w_{tt}$ and $\epsilon _{\alpha\alpha'} = \epsilon _{tt}$ when interactions between two tail beads from different lipid chains are considered. This lipid model was originally constructed to describe a lipid membrane in the absence of solvent~\cite{Cooke_Kremer_Deserno_05}. Since our simulation contains explicit, albeit effective point solvent molecules, the parameters that enter this model were altered (see below) to account for the explicit presence of the solvent molecules.

\subsection{Lipid-solvent interactions}

The solvent particles interact with the lipid beads through intermolecular potentials. The interaction between a solvent particle and a lipid tail bead is also given by Eq.~(\ref{repulsive_potential}) with the same cutoff length $r_c$, but with a different interaction strength $\epsilon _{\alpha\alpha'}=\epsilon_{st}$.  This interaction is purely repulsive; it accounts for hydrophobic effects.  The interaction between a solvent particle and a lipid head bead is given by Eq.~(\ref{attarctive_potential}) with $w_{\alpha\alpha'}=w_{sh}$ and $\epsilon _{\alpha\alpha'}=\epsilon _{sh}$.  This interaction is repulsive at $r<r_c$ and attractive for $r_c <r <r_c+w_{sh}$, so that hydrophilic effects are taken into account.

\subsection{MD-MPC dynamics}

The system consists of $N_L$ lipid molecules and $N_S$ solvent molecules. Since there are no explicit solvent-solvent interactions, the total potential energy of the system, $V_T$, may be written as the sum of interactions within the $N_L$ single lipid molecules, $V_{\ell}$, interactions among different lipid molecules, $V_{\ell \ell}$, and lipid-solvent interactions, $V_{\ell s}$: $V_T=V_{\ell}+V_{\ell \ell}+V_{\ell s}$. Instead of explicit interactions among solvent molecules, their interactions are treated by multiparticle collision dynamics~\cite{Malevantes_Kapral_99}. Hybrid MD-MPC dynamics combines molecular dynamics segments of evolution with effective multiparticle solvent collisions at discrete time intervals $\tau$ to obtain the time evolution of the entire system in the following way\cite{Malevantes_Kapral_00}:

Given that the total potential energy of the entire system is $V_T$, Newton's equations of motion are used to evolve all particles for a time interval $\tau$. Note that because there are no solvent-solvent interactions this MD trajectory segment can be simulated efficiently, even for large systems containing many solvent particles. At time $\tau$ multiparticle collisions among solvent molecules take place. To carry out such collisions, the solvent particles are sorted into the cells of a simple cubic lattice and particles in the same cell exchange momentum with each other while the total momentum in the cell is conserved. We employ the constant temperature version of MPC dynamics~\cite{Gompper_Ihle_Kroll_Winkler_09}. If the mean velocity of the solvent particles in the cell $\xi$ is $\mathbf{V}_{\xi}$, the collision event of the $i$-th particle inside this cell is modeled by updating its velocity, ${\bf v}_i$, so that the new velocity, ${\bf v}'_i$, is given by
\begin{equation}
{\bf v}'_i= \mathbf{V}_\xi + \mathbf{v}_i^{ran}-\sum_{j\in \mbox{cell}\:\xi}\mathbf{v}_j^{ran}/N_{\xi}\mbox{ },
\label{MPC_eq}
\end{equation}
where the components of $\mathbf{v}_i^{ran}$ are chosen as Gaussian random numbers with zero mean and variance $k_BT/m$, $N_{\xi}$ is the number of solvent particles in the cell $\xi$ and the summation is performed over all solvent particles in this cell. Since the mean free path of the solvent particles in our simulation was small compared with the size of a MPC cell, we used random grid-shifting\cite{Ihle_Kroll_01,Ihle_Kroll_03} to implement the MPC step. This sequence of MD and MPC steps is repeated to evolve the entire system. The properties of such MPC dynamics have been discussed in detail in reviews where further applications can be found\cite{Kapral_08,Gompper_Ihle_Kroll_Winkler_09}.

\subsection{Simulation details\label{sec:Numerical_simulation}}

The characteristic interaction energies between different types of beads were $\epsilon_{ht} = 1\:\epsilon$, $\epsilon_{hh}=\:\epsilon_{tt}=0.5\:\epsilon$,  $\epsilon_{sh}=0.05\:\epsilon$ and $\epsilon_{st}=2.0\:\epsilon$. The attraction ranges for tail-tail and solvent-head interactions were chosen such that $r_c+w_{tt}= 2.6\:\sigma$ and $r_c+ w_{sh}= 1.65\:\sigma$.  All particles and beads had equal mass $m$.
The simulations were carried out in a cubic box of size $25\:\sigma \times 25\:\sigma \times 25\:\sigma$ with periodic boundary conditions.
The lateral size of a MPC cell was $a_0=\sigma$. The system contained $1000$ lipid chains and $56624$ solvent particles. On average, the solvent number density in the bulk was equal to five.

The initial velocities of all particles were Gaussian distributed with zero mean and variance $k_BT/m$ for each component.  For the MD trajectory segments, Newton's equations of motion were integrated using the velocity-Verlet algorithm\cite{Rapaprot_04} with a time step of $\delta t=0.005\; t_0$, where $t_0=\sqrt{m\sigma^2/\epsilon}$, and the MPC time step was $\tau=0.2=40\ \delta t$. The initial configuration of the membrane was prepared by arranging the lipids as a bilayer in the $xy$-plane, with the hydrophilic particles facing the outer surfaces while the solvent particles were randomly distributed in the rest of the simulation box. Simulation data was gathered after the system had evolved for $10^5 \;\delta t$, so that thermal equilibrium was established. Depending on the physical quantity under investigation, time averages were taken over time intervals up to $10^6\; \delta t$.
%All simulations were performed using a standard personal computer. A typical run of $10^5\:\delta t$ on a 2.83GHz Intel Core 2 Quad CPU takes about 4 CPU hours.

Results will be reported below in dimensionless simulation units except where connections with physical length and time scales are made. We have chosen $\sigma$ to be the unit of length and $m$ the unit of mass. The characteristic interaction energy between a lipid head and a lipid tail bead, $\epsilon _{ht} \equiv \epsilon$, was taken to be the unit of energy. Time will be reported in units of $\delta t$.

%%%%%%%%%%%%%%%%%%%%%%
\section{Membrane Properties at different temperatures\label{sec:Membrane Properties}}

Self-assembled lipid membranes are known to have a rich phase behavior\cite{Mouritsen_05}. At higher temperatures, the lipids in the membrane are not ordered and the membrane is in the so-called liquid phase.  As the temperature decreases, the membrane undergoes a transition to a gel phase in which the lipid chains show nematic order.  In simulations at three different temperatures, we observed various bilayer structures, which were analyzed by determining the vertical density profile, the lateral radial distribution function, the chain order parameters, and the in-plane diffusion constant of the lipids.

\begin{figure}
\includegraphics[width=1.0\columnwidth]{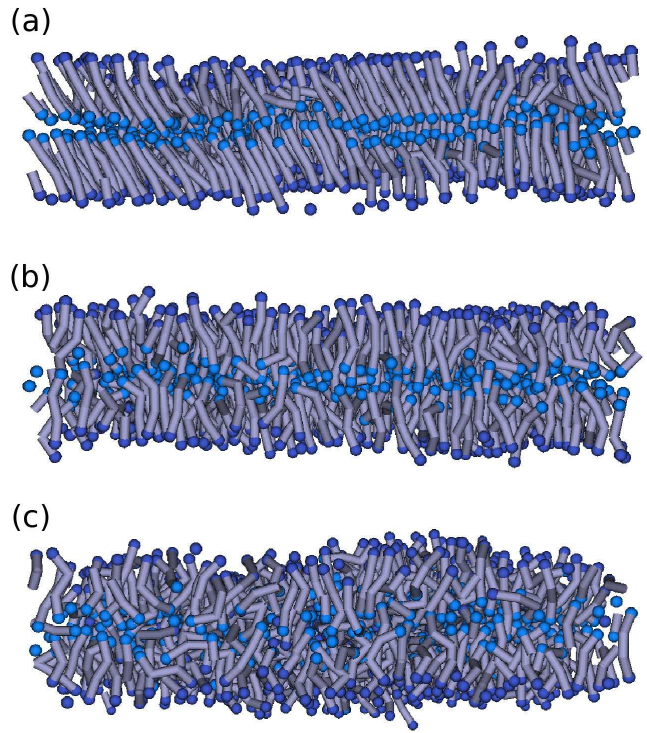}
\caption{\label{fig:membrane_strucutre} Membrane structures at three different temperatures: (a) $k_BT/\epsilon=0.4$, (b) $k_BT/\epsilon=1.0$ and (c) $k_BT/\epsilon=2.0$. The rod representation is used to display lipids. The solvent particles are not shown. }
\end{figure}

Examples of membrane structures observed in our simulations are shown in Fig.~\ref{fig:membrane_strucutre}. We have chosen to visualize the lipids using the rod representation shown in Fig.~\ref{fig:lipid}(b). The FENE bonds are displayed as gray solid rods, with only the hydrophilic head beads (dark blue) and the terminal bead of the hydrophobic tail (light blue) explicitly shown.

At $k_BT/\epsilon=0.4$ (Fig.~\ref{fig:membrane_strucutre}(a)), the lipids are mostly straight and relatively well ordered. Two domains can be seen in the figure. In the majority domain, the lipids are tilted and roughly parallel to one another. In the smaller domain in the right part of the membrane, the lipids are less ordered and the structure is similar to that seen at the higher temperatures. At $k_BT/\epsilon=1.0$, the tilted ordered structure is not observed and the orientation of lipids is less ordered (Fig.~\ref{fig:membrane_strucutre}(b)). Nonetheless, there is still a well-defined  midplane which separates two lipid monolayers. Moreover, it is clear that, on  average, the lipid chains are perpendicular to the midplane of the membrane. When the temperature is further increased to $k_BT/\epsilon=2.0$, an irregular structure is found where the orientational order of the lipid chains is weak and the lipid head particles penetrate into the membrane interior, so that the bilayer midplane and the interface between the lipids and the solvent are less defined (Fig.~\ref{fig:membrane_strucutre}(c)).

To quantitatively characterize the bilayer structures at various temperatures, the orientational order parameter, the in-plane radial distribution function and the vertical density distributions of lipid chains can be used.

The orientational order parameter is defined as $S=\frac{1}{2}\langle3\cos^2\theta_{\ell}-1\rangle$ where the bracket $\langle...\rangle$ denotes a canonical equilibrium average.
For the $\ell$-th lipid chain, $\cos \theta_{\ell} = \hat{\mathbf{r}}_{\ell}\cdot \hat{\mathbf{n}}$, where $\hat{\mathbf{r}}_{\ell}$ is the unit vector pointing from the last tail bead to the lipid head and $\hat{\mathbf{n}}$ is either the unit normal to the upper or lower monolayers. The orientational order of the chains decreases as $S$ diminishes. When $S=1$, all lipid chains are aligned parallel to the bilayer normal. On the other hand, $S=0$ implies that, on average, there is no correlation between the directions of the lipids and the bilayer normal.

In our simulations, the orientational order parameter was determined by averaging over all lipids and over $1000$ bilayer configurations separated by $200\;\delta t$. We found that $S=0.54$ for the membrane at temperature  $k_BT/\epsilon=1.0$, typical for a membrane in the liquid phase\cite{Kranenbury_Venturoli_Smit_03}. At the higher temperature $k_BT/\epsilon=2.0$, the order parameter drops to $S=0.21$, thus indicating a more disordered orientational structure. One might have expected that the lipid chains would have been more ordered at the lower temperature $k_BT/\epsilon=0.4$. However, the orientational order parameter actually decreases to $S=0.46$, since most of the chains are then tilted and therefore their direction deviates from the bilayer normal.

To better characterize chain orientational order at the temperature $k_BT/\epsilon=0.4$, we have chosen a domain where the lipids were tilted and introduced the unit vector $\hat{\mathbf{n}}_t$ pointing along the average direction of the tilted lipids. In this domain, $\cos \theta_{\ell}$ was determined by computing the inner product of $\hat{\mathbf{n}}_t$ and the unit vector $\hat{\mathbf{r}}_{\ell}$ of the lipid. When $\cos \theta_{\ell}$ was defined in this way, we found that $S = 0.96$, confirming that the orientational order of the membrane was even higher at this lower temperature.

\begin{figure}
\includegraphics[width=.8\columnwidth]{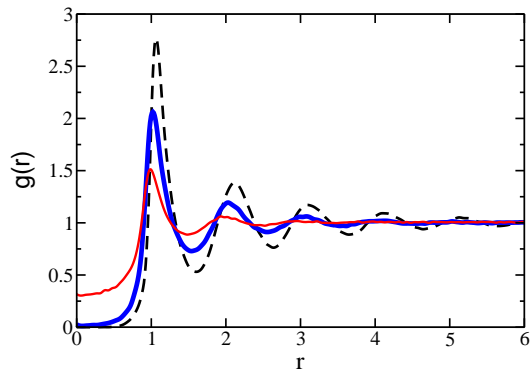}
\caption{\label{fig:RDF} Radial distribution functions of lipid head beads in the membrane at three different temperatures: (a) $k_BT/\epsilon=0.4$ (dashed black line), (b) $k_BT/\epsilon=1.0$ (thick blue line) and (c) $k_BT/\epsilon=2.0$ (thin red line).}
\end{figure}

We have also determined the in-plane radial distribution function $g(r_{\parallel})$ of lipid head beads,
\begin{equation}
g(r_\parallel) = \frac{\rho(r_\parallel|0)}{\bar{\rho}},
\end{equation}
where $\rho(r_\parallel|0)$ is the average two-dimensional density of head beads at a projected distance $r_\parallel$ on the $xy$-plane from a given head bead and $\bar{\rho}$ is the average two-dimensional density of lipid-head beads. To compute this property averages were taken over all lipid head beads in $500$ bilayer configurations separated by $2000 \:\delta t$.
Figure~\ref{fig:RDF} shows radial distribution functions $g(r_\parallel)$ at three different temperatures. When $k_BT/\epsilon = 0.4$, the radial distribution function has several peaks extending to $r_\parallel\approx5$ and the separation between the peaks is close to the size of a lipid bead. At $k_BT/\epsilon=2.0$, the fact that the radial distribution function is not vanishing when $r_\parallel<1$ suggests that the positions of two lipid head beads, projected on the $xy$-plane, overlap due to the presence of lipid head beads in the interior of the membrane. As temperature increases, the in-plane correlations become weaker indicating that the membrane is less structured at higher temperatures.

\begin{figure}
\includegraphics[width=.8\columnwidth]{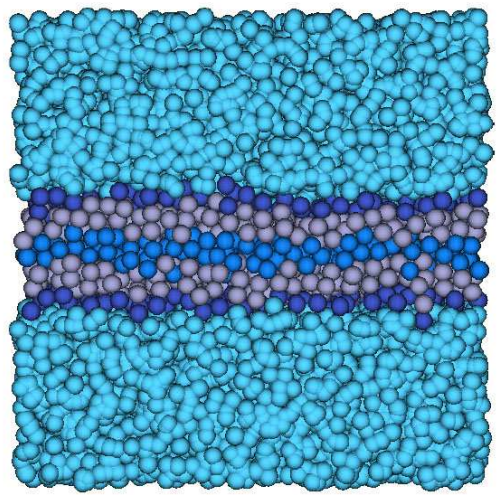}
\caption{\label{fig:cut_of_box} A cut through the simulation box showing the vertical structure of the bilayer and solvent particles at $k_BT/\epsilon=1.0$.}
\end{figure}

\begin{figure}
\includegraphics[width=.8\columnwidth]{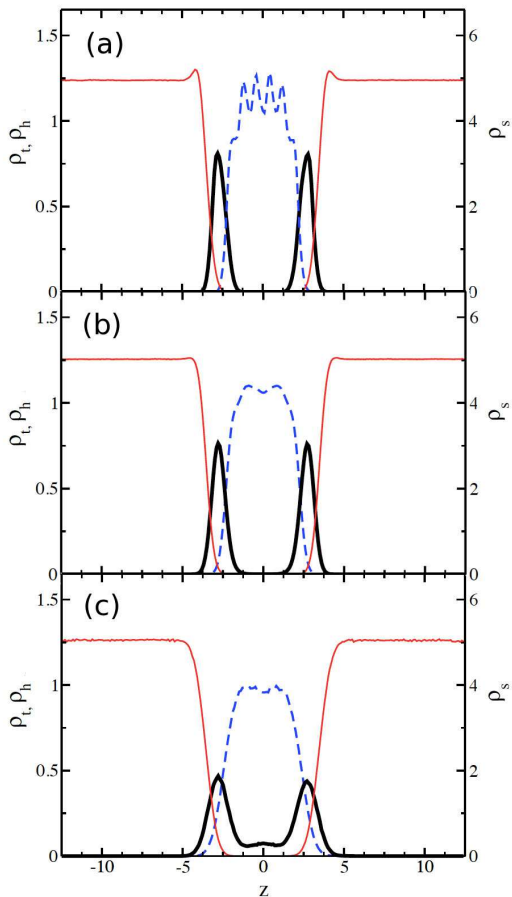}
\caption{\label{fig:density_profile_scale} Vertical density profiles for hydrophilic head beads ($\rho_h$, thick black line), hydrophobic tail beads ($\rho_t$, dashed blue line) and solvent particles ($\rho_s$, thin red line) at three different temperatures : (a) $k_BT/\epsilon=0.4$, (b) $k_BT/\epsilon=1.0$ and (c) $k_BT/\epsilon=2.0$. The scale is different for the solvent density profile.}
\end{figure}

Another important statistical property is the vertical density profile of lipid particles. Figure~\ref{fig:cut_of_box} displays a cut through the simulation box showing the vertical structure of the membrane and surrounding solvent particles. To determine the vertical profiles, the simulation box was divided into $250$ slices in the $z$-direction; each slice had thickness $0.1$. The time-averaged density profiles of solvent ($\rho_s$), lipid head ($\rho_h$) and lipid tail ($\rho_t$) beads for each slice at different temperatures were computed, with the average over all system configurations up to $10^6 \:\delta t$ after the system reached equilibrium. The results are shown in Fig.~\ref{fig:density_profile_scale}.

When $k_BT/\epsilon=0.4$, the density profile of lipid tail beads consists of several sharp peaks, each of which corresponds to the vertical position of one lipid-tail bead, thus indicating a well-ordered vertical arrangement for the beads along a chain and small membrane shape fluctuations (Fig.~\ref{fig:density_profile_scale}(a)). As temperature increases, a smoother profile for the lipid tail density is observed, showing that the beads along a lipid chain are less ordered and thermal  fluctuations of the membrane shape are more significant (Fig.~\ref{fig:density_profile_scale}(b)). At $k_BT/\epsilon=2.0$, one can see that the density of lipid head beads in the interior of the bilayer becomes significant. Moreover, the distribution of lipid tails is also broader and a larger overlap with the distribution of lipid-head beads is observed (Fig.~\ref{fig:density_profile_scale}(c)). These data again indicate a more disordered bilayer structure, close to the onset of membrane dissociation.

The vertical density profile of the lipid beads, the radial distribution function of the lipid-head beads and the lipid chain order parameter give us information on the equilibrium organization of the membrane.  As the equilibrium structure of the membrane changes with temperature, the dynamics of individual lipids should also be affected.

To investigate lipid diffusion, we computed the in-plane mean square displacement of the lipids, MSD$(t)=\langle [\mathbf{r}_{\ell,\parallel}(t) - \mathbf{r}_{\ell,\parallel}(0)]^2\rangle$, where $\mathbf{r}_{\ell,\parallel}$ is the position of the center of mass of a lipid projected on the $xy$-plane. The averages were taken over all lipids at every $1000\:\delta t$ and the MSD was computed from system trajectories of length up to $10^5\:\delta t$. Depending on the time domain, both diffusive and subdiffusive types of behavior of lipids were found.

\begin{figure}
\includegraphics[width=.8\columnwidth]{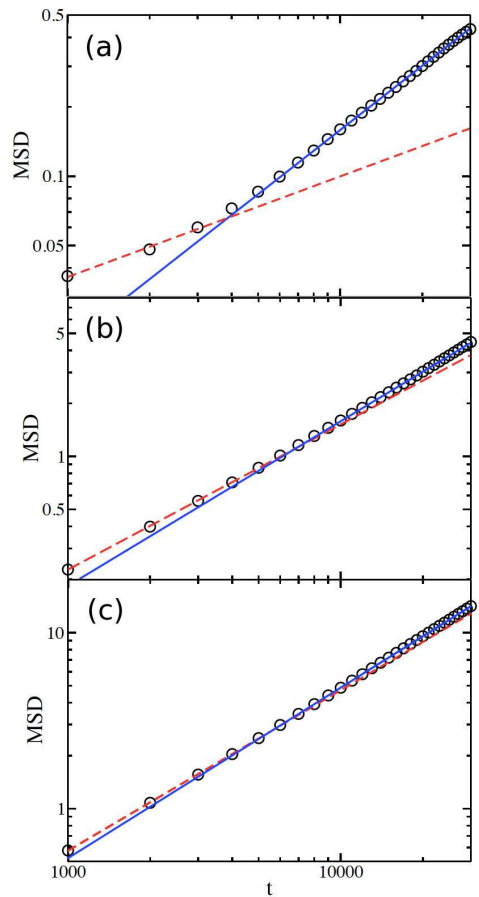}
\caption{\label{fig:diffusion}Diffusion of lipids in the membrane. Log-log plots of the mean square displacements (MSD) of the center of mass of a lipid are shown as functions of time for (a) $k_BT/\epsilon=0.4$, (b) $k_BT/\epsilon=1.0$ and (c) $k_BT/\epsilon=2.0$. The dashed and solid straight lines are linear fits for the subdiffusive and normal diffusive regimes, respectively.}
\end{figure}

For all three temperatures lipid diffusive motion was always observed in the long-time regime, so that MSD$(t) =4Dt$, where $D$ is the diffusion constant. At $k_BT/\epsilon=0.4$ (Fig.~\ref{fig:diffusion}(a)), the diffusion constant is $D=7.87\times10^{-6}$, implying that a lipid moved over a distance approximately equal to the size of a coarse-grained lipid-head bead within the simulation time of $t\sim 10^6\:\delta t$. At a higher temperature $k_BT/\epsilon=1.0$ (Fig.~\ref{fig:diffusion}(b)), the diffusion constant is $D=7.86\times10^{-5}$, which is about $10$ times larger than that at the lower temperature. The diffusion constant further increases to $D=1.86\times10^{-4}$ at $k_BT/\epsilon=2.0$ (Fig.~\ref{fig:diffusion}(c)). Thus, lipid diffusion in the bilayer depends strongly on temperature.

In addition to normal lipid diffusive dynamics in the long-time limit, subdiffusive motion was found at intermediate times, so that MSD$(t)\sim t^{\alpha}$, where $\alpha < 1$ is the subdiffusive exponent. At $k_BT/\epsilon=0.4$ (Fig.~\ref{fig:diffusion}(a)), we found $\alpha={0.44}$ over the time up to the crossover time $t_c \approx 3900\:\delta t$. As the temperature increases, the subdiffusion exponent $\alpha$ grows to $0.82$ and $0.9$ for $k_BT/\epsilon=1.0$ and $k_BT/\epsilon=2.0$, respectively. Similar subdiffusive behavior has been recently observed in all-atom MD simulations of lipid membranes\cite{Akimoto_Yamamoto_Yasuoka_Hirano_Yasui_11}, where in-plane motions of lipids were subdiffusive on the time scale of nanoseconds.

Our simulation data suggests that the membrane at $k_BT/\epsilon=1.0$ is in liquid phase, with a well-defined solvent-lipid interface and bilayer midplane. Since liquid-like membrane states are most relevant for real biological membranes, all our subsequent simulations reported below were carried out at $k_BT/\epsilon=1.0$.

At this temperature, the mean vertical distance between the peaks in the lipid head density profiles for two monolayers (see Fig.~\ref{fig:density_profile_scale}(b)), which can be chosen as the mean thickness of the membrane, is close to $5.5\:\sigma$. Comparing this to the thickness of a typical real membrane (between $4$ and $6$ nm), we can identify the unit length in our simulations to be $\sigma \simeq 1$ nm.  The time unit $\delta t$ can be roughly estimated by comparing the typical lipid lateral diffusion constant in experiment\cite{Korlach_99} ($\approx 4\:\mu \mbox{m}^2$/s) with our computed values. We find $\delta t \simeq 20$ ps. Hence, the physical size of the simulation box is about $25$ nm and the total simulation time is about $20\:\mu$s for the simulation with $10^6\:\delta t$.

\section{Self-assembly of the membrane\label{sec:self_assembly}}
To verify that the lipid bilayer is indeed thermodynamically stable at $k_BT/\epsilon=1.0$, simulations of the membrane self-assembly process at that temperature were performed. The initial condition was taken to be a random mixture of lipids and solvent particles. To prepare this initial configuration, the following procedure was employed: starting with an equilibrium lipid bilayer, the attractive potential interactions between the lipid tail beads were switched off. In addition, attractive interaction potentials between the tail beads and solvent particles, with $\epsilon_{st}=0.1$ and $r_c+w_{st}=2.6$, were introduced, making lipids more hydrophilic. After $2\times 10^5\:\delta t$, the lipid and solvent particles were found to be uniformly distributed within the simulation box.

\begin{figure}
\includegraphics[width=1.0\columnwidth]{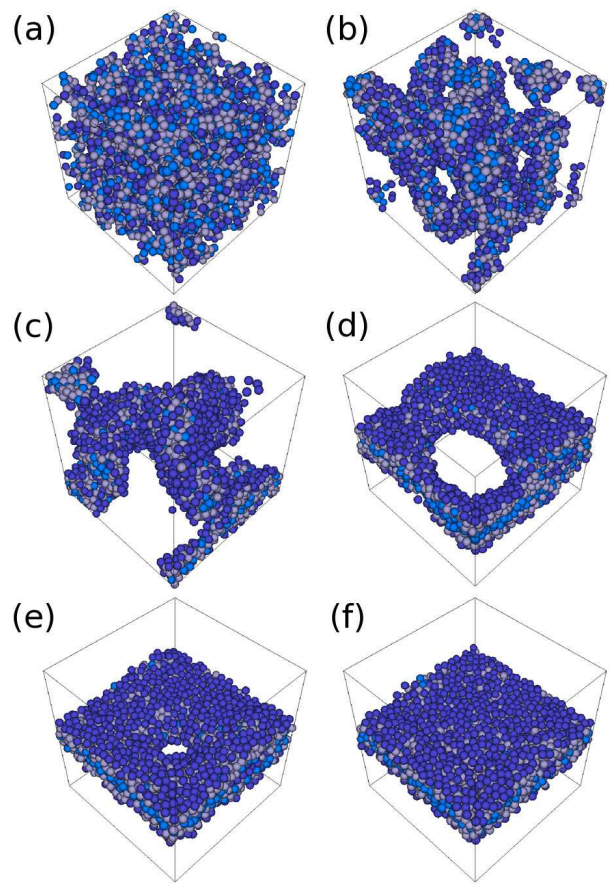}
\caption{\label{fig:snapshots_shift}Self-assembly of the lipid bilayer. Six configurations at time moments (a) 0 $\delta t$, (b) 6000 $\delta t$, (c) 18000 $\delta t$, (d) 160000 $\delta t$, (e) 246400 $\delta t$ and (f) 300000 $\delta t$ are shown. The initial state (a) corresponds to the uniform mixture of lipids and the solvent. (Enhanced online)}
\end{figure}

Starting from this uniform initial configuration, simulations with the full potential model were performed. Figure~\ref{fig:snapshots_shift}(a) shows the initial configuration at $t=0$ where the lipid chains are uniformly distributed inside the simulation box. As time evolves, the lipids quickly aggregate forming small segments (Fig.~\ref{fig:snapshots_shift}(b)). These small segments gradually merge into large branched clusters (Fig.~\ref{fig:snapshots_shift}(c)). Then, a slow rearrangement process takes place leading to a single bilayer structure with a large hole in the center (Fig~\ref{fig:snapshots_shift}(d)). Later, this large hole slowly shrinks to a small pore (Fig~\ref{fig:snapshots_shift}(e)) and eventually closes after $3\times 10^5\:\delta t$ (Fig~\ref{fig:snapshots_shift}(f)). The formation of the lipid bilayer and the close-up of the membrane pore suggest that the uniform, flat membrane is indeed a thermodynamically stable structure at $k_BT/\epsilon=1.0$. (See on-line video of the evolution process.)

\section{Macroscopic membrane properties and velocity correlation functions\label{sec:Statistical_properties}}
Macroscopic theories of biomembranes are formulated in terms of elastic deformable surfaces. There are also hydrodynamic descriptions which treat the membranes as two-dimensional fluids immersed in the three-dimensional solvent. Below we show that our simulation data is consistent with such macroscopic theories. Moreover, our mesoscopic simulations allow us to determine some of the characteristic properties of the lipid bilayer membrane. The analysis is restricted to the liquid-phase membrane at $k_BT/\epsilon$ = 1.0.

The surface tension of the membrane can be obtained by considering the membrane stretching energy. In the regime where Hookian elasticity theory holds, the stretching elastic energy of a membrane with area $A$ is given by
\begin{equation}
E_{s}=\frac{1}{2}K_A \frac{(A-A_0)^2}{A_0},
\label{eq:stretching_energy}
\end{equation}
where $K_A$ is the stretching modulus and $A_0$ is the reference area corresponding to a tensionless membrane. The membrane surface tension is $\gamma = \mbox{d} E_s/\mbox{d}A$ and therefore the equation
\begin{equation}
\gamma =K_A\frac{A-A_0}{A_0}
\label{eq:surface_tension_vs_area}
\end{equation}
holds. In our simulations, we determined the equilibrium surface tension of the membranes with different areas confined to boxes of different lateral sizes. From these measurements, the stretching modulus could also be obtained.

The surface tension of a membrane was determined in our simulations from the pressure tensor using the relation\cite{Jakobsen_05},
\begin{equation}
\gamma = \bigg\langle L_z\bigg[P_{zz} - \frac{1}{2} (P_{xx}+P_{yy}) \bigg] \bigg\rangle,
\label{eq:surface_tension}
\end{equation}
where the bracket denotes an equilibrium canonical average and $L_z$ is the linear size of the simulation box in the $z$-direction. The diagonal elements of the pressure tensor are defined as
\begin{equation}
P_{\nu \nu} = \frac{1}{V} \bigg[ \sum_i v_{i,\nu}v_{i,\nu}+ \frac{1}{2} \sum_{i\neq j} r_{ij,\nu} f_{ij,\nu}\bigg ], \mbox{ } \nu = x,y,z,
\end{equation}
where $V$ is the volume of the simulation box. The summations in this equation are taken over all particles, including both the solvent and the lipids. The $\nu$-component of the distance between two particles, $i$ and $j$, is $r_{ij,\nu}$ and the force acting between them is $f_{ij,\nu}$. When both particles are solvent, $f_{ij,\nu}=0$; otherwise, $f_{ij,\nu}$ is evaluated through the actual interaction potentials. The equilibrium average in Eq.~(\ref{eq:surface_tension}) was computed by a time average over an interval of $10^6 \:\delta t$.

\begin{figure}
\includegraphics[scale=0.27]{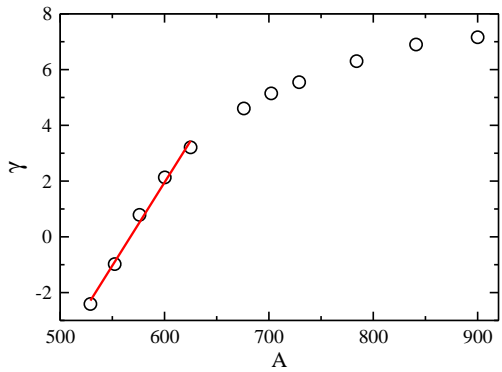}
\caption{\label{fig:area_surface_tension}Dependence of the surface tension $\gamma$ on the membrane area $A$. The first five data points were used to determine the membrane stretching modulus.}
\end{figure}

The computed values of the surface tension for different membrane areas are shown in Fig.~\ref{fig:area_surface_tension}. When the surface tension is small, it depends approximately linearly on the area $A$. By fitting this linear dependence to Eq.~(\ref{eq:surface_tension_vs_area}), we determined the stretching modulus $K_A$ and the area $A_0$ corresponding to the tensionless membrane. We found that $A_0\simeq567\mbox{ }\sigma^2$ and $K_A\simeq33.9\mbox{ }\epsilon/\sigma^2$, In Sec.~\ref{sec:Membrane Properties} we noted that $\sigma \simeq 1$ nm. Since our simulations were performed at $k_BT/\epsilon=1.0$, we have $\epsilon = k_BT$. Therefore, the computed stretching modulus is approximately $K_A=33.9\mbox{ }k_BT/\mbox{nm}^2$. This is comparable to the values observed for typical liquid-like membranes \cite{Phillips_Kondev_Theriot_09}, i.e. $K_A=50-70\:k_BT/$nm$^2$.

In the macroscopic continuous approach\cite{nelson04}, the Helfrich free energy of the membrane is
\begin{equation}
\begin{aligned}
F =& \mbox{ }\frac{1}{2}\int_{L\times L} d^2\mathbf{r}\mbox{ }[\kappa (\bigtriangledown^2 h(\mathbf{r}))^2 + \gamma(\bigtriangledown h(\mathbf{r}))^2]
\\
=& \mbox{ }\frac{1}{2}\sum_{\mathbf{q}}L^2|h_{q}|^2[\kappa q^4 + \gamma q^2]\mbox{ }\mbox{ },
\end{aligned}
\end{equation}
where $h(\mathbf{r}) = \sum_{\mathbf{q}} h_{q} e^{i\mathbf{q}\cdot \mathbf{r}}$ is the local height of the membrane measured with respect to the reference plane. As implied by the energy equipartition theorem, the power spectrum of membrane height fluctuations should therefore be
\begin{equation}
S(q) \equiv L^2\langle |h_q^2|\rangle= \frac{k_BT}{\kappa q^4+\gamma q^2}\mbox{ },
\label{eq:power_spectrum}
\end{equation}
where $\kappa$ is the membrane bending modulus and $\gamma$ is again the surface tension. There exists a characteristic wavenumber $q_c=\sqrt{\gamma/\kappa}$ separating two different regimes. When $q\ll q_c$, the power spectrum is $S(q) \sim q^{-2}$ and the dominant contribution comes from the membrane tension. For $q\gg q_c$, the power spectrum is $S(q) \sim q^{-4}$ and the dominant role is played by the bending elasticity.

To determine membrane height fluctuations, a bilayer configuration from a simulation was taken at every $5000 \:\delta t$ so that, in total, $1000$ bilayer configurations were recorded. In each bilayer configuration, the membrane was divided into a grid of $10\times10$ cells. The membrane height of each cell was further determined by taking the average of the positions of  end beads in the hydrophobic lipid tails. In this way, local heights could be determined at all grid points. Performing a fast Fourier transform for the membrane heights, the power spectrum $S(q)$ could be determined for each bilayer configuration. By averaging over all $1000$ bilayer configurations, the mean power spectrum was obtained. Note that, based on our simulations, the power spectrum could only be computed in the range $q_{max}<q<q_{min}$. Here $q_{max}=2\pi/l$ is twice the linear size of a grid cell, close to the bilayer thickness, and $q_{min}=2\pi/L$, where $L$ is the linear dimension of the simulation box. In our simulations, we had $L=25\:\sigma$ and $l=5\:\sigma$, so that $q_{max}\simeq1.25\:\sigma^{-1}$ and $q_{min}\simeq0.25\:\sigma^{-1}$.

\begin{figure}
\includegraphics[width=.8\columnwidth]{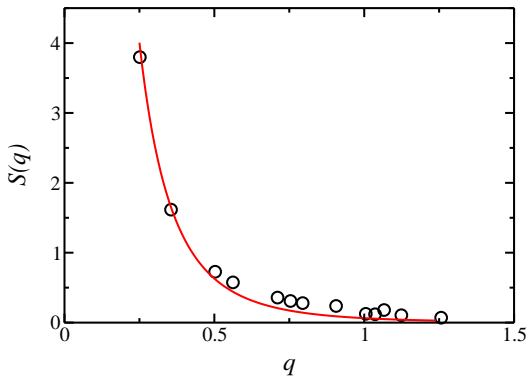}
\caption{\label{fig:PS_membrane}Power spectrum $S(q)$ of membrane height fluctuations. The solid line is the best fit of the simulation data, using the theoretical dependence (Eq.~(\ref{eq:power_spectrum}) ).}
\end{figure}

Figure~\ref{fig:PS_membrane} displays the numerically determined power spectrum. The solid line shown in Fig.~\ref{fig:PS_membrane} is obtained by least-squares fitting using Eq.~(\ref{eq:power_spectrum}) with the membrane tension value $\gamma=3.21\:k_BT/$nm$^2$ taken from the constant surface area simulations (Fig.~\ref{fig:area_surface_tension}). As a result of data fitting, the membrane bending modulus was found to be $\kappa \simeq 12.56\: \epsilon \simeq 12.56\:k_BT$. Using this value of the bending modulus $\kappa$ and the previously determined value of the surface tension $\gamma$ for the membrane, the characteristic wavenumber $q_c = 0.5\:\sigma^{-1}$ could be obtained. This wave number lies in the middle of the computed power spectrum, indicating that our simulations are able to reproduce both the tension-dominated and the bending-dominated regimes. Typical experimental values of the bending modulus $\kappa$ for lipid membranes lie \cite{Phillips_Kondev_Theriot_09} between $10$ and $20\:k_BT$. Hence, we can again notice that the membranes in our simulations are similar in their physical properties to real biological membranes.

Finally, we consider flow dynamics of lipids in the membrane.  In the classical study by Saffman and Delbr\"{u}ck  \cite{Saffman_Delbruck_75}, the membrane was treated as a two-dimensional (2D) simple fluid embedded in a three-dimensional (3D) solvent. When a lipid moves in the membrane, its momentum may be transferred not only to the neighboring lipids, but also to the solvent. However, estimates show \cite{Saffman_Delbruck_75,Diamant_09} that, on length scales shorter than a micrometer, hydrodynamic coupling between the membrane and the solvent is not significant and, on such scales, the membrane can be approximately treated as a 2D fluid.

The longitudinal and transverse velocity correlation functions of lipid flows are
\begin{equation}
\begin{aligned}
C_L(x,t) &= \big\langle v_x(x_0,y_0,t_0)v_x(x_0+x,y_0,t_0+t) \big\rangle_{x_0,y_0,t_0},
\\
C_T(x,t) &= \big\langle v_y(x_0,y_0,t_0)v_y(x_0+x,y_0,t_0+t) \big\rangle_{x_0,y_0,t_0}.
\end{aligned}
\label{eq:velocity_corre_1}
\end{equation}
where angular bracket $\langle \dots  \rangle_{x_0,y_0,t_0}$ denotes an average over the positions $x_0$, $y_0$, time $t_0$ and realizations. The hydrodynamic velocity field $\mathbf{v}(x,y,t)$ is defined by taking the average of the instantaneous velocities of all lipids within a certain membrane area element. In our simulations, the membrane was divided into a grid of $10 \times 10$ of cells and the hydrodynamic velocities were obtained by averaging the in-plane lipid velocities in each cell. The  products $v_x(x_0,y_0,t_0)v_x(x_0+x,y_0,t_0+t)$ and $v_y(x_0,y_0,t_0)v_y(x_0+x,y_0,t_0+t)$ were determined for all grid points $(x_0, y_0)$ at every MD step, and the correlation functions $C_L(x,t)$ and $C_T(x,t)$ were computed by taking the average of these products over all grid points $(x_0,y_0)$ and over $1000\:\delta t$. Subsequently, the results were additionally averaged over an ensemble of $20$ independent realizations.

\begin{figure}
\includegraphics[width=.8\columnwidth]{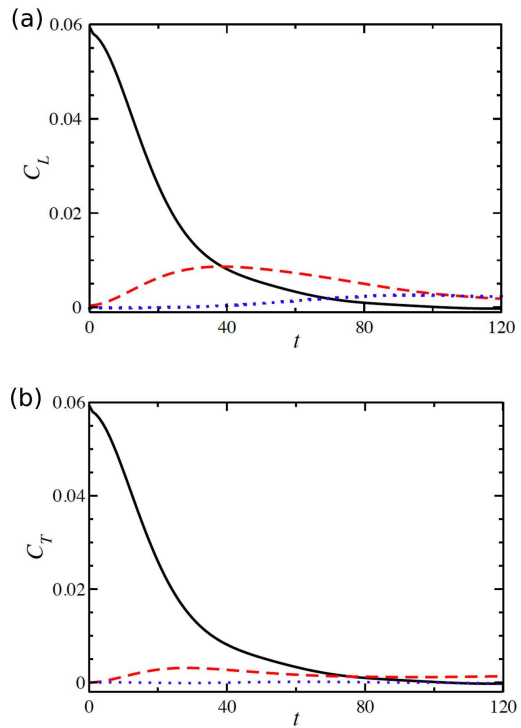}
\caption{\label{fig:v_correlation_dx_dy} Time dependence of the longitudinal (a) and transverse (b) velocity correlation functions for three different separations: $x=0$ (solid lines), $x=0.25$ (dashed lines), and $x=5$ (dotted lines).}
\end{figure}

Figure~\ref{fig:v_correlation_dx_dy} shows the dependences of $C_L(x,t)$ and $C_T(x,t)$ on time for three different values of the distance $x$. The peak in $C_L(2.5,t)$ is found $40\:\delta t$ later than the peak in $C_L(0,t)$, suggesting that it takes $40\:\delta t$ for a fluctuation of velocity $v_x$ to be transported over a distance $x=2.5$. Similarly, it takes about $25\:\delta t$ for the fluctuations of $v_y$ to be transported in the $x$-direction over such distances. This is much faster than the time, $x^2/D \sim 8\times 10^4\:\delta t$ for $x=2.5$, needed for the lipids to diffuse over the same distance. Therefore, we conclude that velocity fluctuations are transported by collective lipid flows, not by the diffusion of single lipids.

We can also consider the time integrals of the velocity correlation functions,
\begin{equation}
\begin{aligned}
G_L(x) &= \int_0^{\infty} C_L(x,t)\mbox{ }dt,
\\
G_T(x) &= \int_0^{\infty} C_T(x,t)\mbox{ }dt.
\end{aligned}
\label{eq:velocity_corre_2}
\end{equation}
They are determined by the pair mobility tensor which describes the velocity response of one fluid element due to the motion of another element in the fluid~\cite{Diamant_09}. Such responses are given by the Green function of the Stokes equation. The behavior of the Green functions depends on the dimensionality of the fluid. For three-dimensional fluids, the functions fall as $1/r$ with the distance $r$. In contrast to this, logarithmic distance dependence is characteristic for two-dimensional fluids.

As suggested by Saffman and Delbr{\"u}ck \cite{Saffman_Delbruck_75}, biomembranes can be viewed as 2D fluids of lipids which are immersed in a 3D solvent. On length scales typical for our simulations, viscous coupling between the membrane and the solvent is negligible. Assuming that the membrane is a planar 2D fluid, expressions for the longitudinal and transverse velocity fluctuations can be derived from the pair mobility tensor \cite{Diamant_09}. Thus, one gets
\begin{equation}
\begin{aligned}
G_L(x) &= -C\ln (x/R_c),
\\
G_T(x) &= -C \big[\mbox{ }1+ \ln (x/R_c) \mbox{ } \big],
\end{aligned}
\label{eq:velocity_corre_3}
\end{equation}
where $C$ is a constant prefactor and $R_c$ is a cutoff length which is typically on the micron scale. These approximate expressions hold for distances $x < R_c$. On longer length scales,  momentum diffusion into the bulk solvent becomes significant and a crossover to the behavior characteristic for 3D systems should take place. Note that for a finite system, $R_c$ should be approximately equal to the linear system size \cite{Diamant_09}.

\begin{figure}
\includegraphics[width=.8\columnwidth]{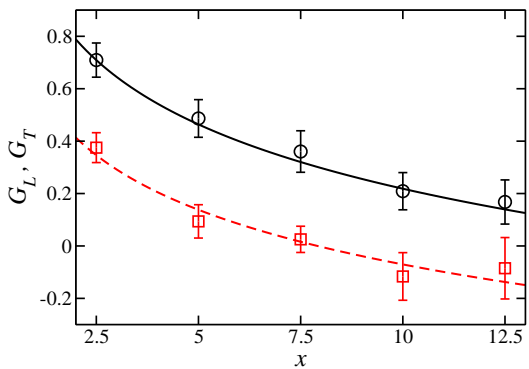}
\caption{\label{fig:v_correlation}Longitudinal (circles) and transverse (squares) correlation functions $G_L(x)$ and $G_T(y)$. The solid and dashed lines show the respective logarithm approximations given by Eq.~(\ref{eq:velocity_corre_3}).}
\end{figure}

Figure~\ref{fig:v_correlation} displays $G_L(x)$ and $G_T(x)$, the longitudinal and transverse correlation functions, determined in our simulations. The solid and dashed lines show best fits using the logarithmic approximations (\ref{eq:velocity_corre_3}) with $C\simeq1.6\times 10^{-3}$ and $R_c\simeq20$. Good agreement is found indicating that the lipid flows in our simulations were indeed well described in terms of 2D hydrodynamics and that the leakage of lipid momentum into the solvent was negligible on the length scale of our system.

\section{Discussion and Conclusions \label{sec:Conclusion}}
We have presented and tested a coarse-grain simulation method for biomembranes. In common with other coarse-grain methods, individual lipids were modeled as short chains of beads linked by elastic bonds, and the solvent was explicitly included using multiparticle collision dynamics.  Our method differs from other investigations \cite{Noguchi_Gommpper_05,Inoue_Takagi_Matsumoto_08} of lipid membrane where MPC dynamics for the solvent was employed in that we account both for the structure of lipid bilayer and include explicit lipid-solvent hydrophobic and hydrophilic interactions.

The interaction parameters of the model were chosen to reproduce the behavior of typical real lipid bilayers. Thus, we could follow in our simulations the self-assembly of a membrane starting from a uniform mixture of lipids and solvent. We could also reproduce various  structural states of lipid bilayers at different temperatures, including the gel phase at the lower temperature and the liquid phase at the higher temperature. 
 
Statistical properties of collective modes of the liquid state of the membrane were studied. By varying the membrane area, the membrane surface tension was determined and the lateral stretching modulus were obtained. The bending modulus of the membrane was then derived from the power spectrum of membrane height fluctuations.  The results show that the elastic properties of our model membranes are comparable to those of a typical real lipid bilayer.

Hydrodynamics of membrane flows was numerically investigated by computing correlation functions of the lipid velocity field.  We found that the velocity fluctuations are not due to the diffusion of single lipids but are propagated by collective hydrodynamic modes. The computed velocity correlation functions show logarithmic spatial dependence, suggesting that, on the length scale of our simulations, the lipid bilayer could be considered as a 2D viscous fluid with little momentum diffusion into the bulk solvent.

Our simulation method has a number of advantages. By modeling the solvent using multiparticle collision dynamics, one does not need to expend computational power to calculate forces acting between solvent particles, as in MD and DPD simulations. Thus, the simulations could be substantially accelerated. 

Another important feature in our simulations was that the lipid-lipid and lipid-solvent interactions both contained short-range hardcore repulsion.  Therefore, crowding effects in the lipid membrane could be well reproduced, as seen in the observed short-time subdiffusive motion of single lipid chains.  This effect was previously reported in an all-atom MD study \cite{Akimoto_Yamamoto_Yasuoka_Hirano_Yasui_11}, but the long-time normal diffusive regime of single lipid chains was not found.

Finally, we would like to point out that it is possible to combine our fast coarse-grain descriptions of membranes and solvent with coarse-grain simulations for proteins \cite{Cressman_Togashi_Mikhailov_Kapral_08}. Such structurally-resolved numerical investigations of individual protein machines in biomembranes, as well as the collective dynamics of such protein machines, will be presented in future work.

Financial support from the Humboldt Foundation and the DFG Training Research Group (GRK 1558) ``Nonequilibrium collective dynamics in condensed matter and biological systems" in Germany is gratefully acknowledged. The research of RK is supported in part by the Natural Sciences and Engineering Research Council of Canada. The research of MJH an HYC is supported by the National Science Council of the Republic of China (Taiwan) under Grant No. NSC 98-2112-M-008-004-MY3 and by the National Center for Theoretical Sciences, Taiwan.

\bibliography{reference}

\end{document}